\documentclass[aps,preprint]{revtex4-2}

\usepackage{lineno}
\usepackage{dcolumn}

\usepackage{bm}
\usepackage{amsmath}
\usepackage{multirow}
\usepackage{graphicx}
\usepackage{float}
\usepackage[colorlinks=true]{hyperref}

\newcommand{\be}{\begin{equation}}
 \newcommand{\ee}{\end{equation}}

\begin{document}

\title{Reply to Comment on ``Nonlinear quantum effects in electromagnetic radiation of a vortex electron'' \\ by Aviv Karnieli, Roei Remez, Ido Kaminer, et al.}

\author{D.\,V.~Karlovets}


\affiliation{School of Physics and Engineering, ITMO University, 197101 Saint-Petersburg, Russia}

\author{A.\,M.~Pupasov-Maksimov}


\affiliation{Universidade Federal de Juiz de Fora, Brasil} 




\date{\today}

\begin{abstract}
{\small We argue that while the experiment of Remez et al. [PRL {\bf 123}, 060401 (2019)] is interesting and its conclusions may well be correct, the observed lack of dependence of the measured angular distributions on the electron's transverse coherence length should have been expected for the parameters chosen. This is because for Smith-Purcell radiation it is the coherence length of a virtual photon $\sigma_{\perp}^{(\gamma)} \approx \beta\gamma\lambda \lesssim \lambda$ that plays a role of the radiation formation width and not the entire electron's coherence length that can well be orders of magnitude larger than the former. This is a common feature for all the radiation processes in which a photon is emitted not directly by the electron packet, which can be delocalized in space, but rather by a much better localized atom or a conduction electron on a surface. Therefore, in our opinion the results of Remez et al. cannot rule out the alternative hypothesis of the delocalized charge. The question, mainly addressed in the comment by Karnieli et al., of whether the measurements were performed in the wave zone or not is interesting but \textit{secondary}. We emphasize that the measured azimuthal distributions are \textit{unusually wide} and neither the original paper nor the recent comment fully discusses and rules out all alternative classical hypotheses that could have led to the same wide distributions. On the contrary, there exists a family of classical effects that could also have resulted in the measured distributions and that were neither discussed nor even mentioned by the authors. Such alternative hypotheses include {\it (i)} effects of the beam sizes, of its angular divergence, of the temporal coherence of the radiation process, which is also related to how the wave zone is defined, and {\it (ii)} influence of the grating shape and of its material -- the effects that are known to be of crucial importance for Smith-Purcell radiation from non-relativistic electrons. 
Finally, we propose to repeat the experiment and to measure diffraction radiation from a thin metallic semi-plane (or a strip) or Compton emission in a laser pulse. 
In these cases, the aforementioned classical effects play a much smaller role and the results of such measurements would have higher credibility.}
\end{abstract}

\maketitle

\section{Introduction}

In their comment \cite{com} to our paper \cite{PRA} Karnieli et al. argue that
\begin{itemize}
\item
Our paper is unable to explain the experiment \cite{exp} (the end of page 1 in \cite{com});
\item
We claim ``that a semiclassical interpretation could still explain the physics of the experiment
reported in Ref.\cite{exp}... The main reason ... is that the experiment in Ref.\cite{exp} was performed in near-field
conditions.''
\item
The calculations in our work \cite{PRA} ``only hold for a certain experimental
situation where the final state of the emitting electron is observed in coincidence with the
emitted photon''.
\end{itemize}

We argue that while two former claims are based on the distortion and misinterpretation of our views reported in Ref.\cite{PRA}, the latter claim is probably due to inattentive reading of our paper.
Our central idea (see page 2 in \cite{PRA}) is that the main result of Ref.\cite{exp}, which is the lack of dependence of the azimuthal distributions on the electron's transverse coherence length, 
could have been expected for the parameters chosen and no hypothesis of the localized nature of the electron's charge is needed to explain it. This is because, in contrast to emission in external fields, 
a photon in the Smith-Purcell effect is emitted not by the electron packet itself, which can be quite wide, but by an atom or a conduction electron on a grating's surface, which are much better localized in space. 
This is also the case for such processes as, say, transition radiation or diffraction radiation. Regardless of how wide the electron packet is, the radiation is due to scattering of a virtual photon by an atom,
the transverse coherence length of which is $\sigma_{\perp}^{(\gamma)} \approx \beta\gamma\lambda \lesssim \lambda$ for $\beta = u/c \approx 0.4-0.7$ and $\gamma = 1/\sqrt{1-\beta^2}$.

Thus no near field is needed to explain the main result of Ref.\cite{exp}, which is why the question, addressed in the comment by Karnieli et al., of whether the measurements were performed in the wave zone or not is interesting but \textit{secondary}. If the measurements \textit{were} in the wave zone, one could still expect the obtained lack of dependence on the transverse coherence merely based on the above physical picture of the emission process but this \textit{does not} allow one to conclude in favor of one of the hypotheses discussed in Ref.\cite{exp}. Next, we stress that the azimuthal distributions reported in Ref.\cite{exp} are \textit{unusually wide} (several well-known models of the Smith-Purcell radiation predict much narrower distributions) and this fact \textit{remains unexplained} by the authors of the comment. 

Moreover, in their comment by Karnieli et al. \cite{com} and in the original paper by Remez et al. \cite{exp} the authors call the semiclassical interpretation an approach in which the probability current density ${\bm j} \propto {\bm u} |\psi|^2$ for an electron with a wave function $\psi$ is interpreted as a source of radiation in the right-hand side of the Maxwell equations. In our opinion, such an approach is initially incorrect because the probability current itself does not represent an observable and, therefore, it cannot enter the Maxwell equations. So the dichotomy discussed by the authors looks rather artificial from a calculational point of view.
On the other hand, we emphasize that the very idea of the experiment \cite{exp} to compare emission patterns for electrons of the different spatial coherences looks very promising. 
However in our view it should be tested in the problems in which it is the electron itself that emits a photon such as, for instance, Compton emission in a laser wave. 

Despite not being directly relevant to the main topic of this discussion, below we further elucidate that the radiation formation width and the corresponding radius of the wave zone
are also connected with \textit{the temporal coherence} of the radiation process. We provide the necessary quantitative estimates of a time interval during which 
the radiation is formed and argue that the wave zone is defined by the transverse coherence length only for radiation of a single electron,
whereas a finite current of many electrons results in \textit{two-photon correlations}, which can contribute to the measurements for large statistics and which represent a source of systematic uncertainties.
The authors of the experiment \cite{exp} and of the comment \cite{com} do not discuss the errors of their measurements and their one-particle theory is not applicable to account for these collective effects.

Furthermore, there are several other classical effects that have not been mentioned in Refs.\cite{exp, com} and that can also result in broadening of the azimuthal distributions,
thus mimicking the quantum effect of the spatially localized charge. They include the role of the grating shape, of its material, and of the beam angular divergence. 
For Smith-Purcell radiation, these effects are known to be of secondary importance only for ultrarelativistic energies, $\beta = u/c \approx 1, \gamma = 1/\sqrt{1 - \beta^2} \gg 1$, but for parameters of Ref.\cite{exp}, $\beta \approx 0.7$, they can well lead to the observed broadening.

Finally, we note that the claim that our results in \cite{PRA} “only hold for a certain experimental situation where the final state of the emitting electron is observed in coincidence with the emitted photon” is incorrect and is likely based on inattentive reading of our paper. We actually studied two scenarios: when the final electron is observed in coincidence with the final photon and when it is not detected at all.
We further elucidate that the undetected final electron should be described as a plane wave with a definite momentum, \textit{as if it were in fact detected} by an ordinary plane-wave detector,
because the plane-wave electron is completely delocalized in space and its momentum simply follows from the energy-momentum conservation law, provided that the final photon is detected with a certain frequency 
and the grating's recoil can be neglected. The situation here is somewhat analogous to that in the theory of neutrino oscillations in which the undetected neutrino is also described as a plane wave \cite{Akh}.

\section{Pre-wave zone and temporal coherence}

The pre-wave zone effect arises because of \textit{partially destructive interference} between the waves emitted from a source of a finite spatial extent.
For radiation from a single electron, the width where the radiation is formed is the transverse coherence length of the electron packet. When several electrons can emit several photons \textit{nearly simultaneously}, these photons can interfere while propagating to the detector. As a result, the whole beam width defines the region of the radiation formation, not only the transverse coherence length of single electrons, and the pre-wave zone radius becomes larger. 

Unfortunately, neither the paper \cite{exp} nor the comment \cite{com} contains a quantitative justification for neglecting these collective effects. As can be seen in Fig.3 a,b of Ref.\cite{exp}, 
the radiation formation length is much larger than the interaction length and is several times larger than the average distance between electrons in the beam. So the generation of at least two photons 
nearly simultaneously by different electrons with a random transverse shift can happen and one needs to estimate the contribution of such a process. The simulations in Refs.\cite{exp,com} were performed 
for a single electron emitting a single photon and the intensities were summed incoherently. For radiation from a beam, such an approach is applicable only in the wave zone 
where interference between the waves emitted by different electrons is constructive, while it is \textit{inapplicable} in the pre-wave zone. For the low current used in Ref.\cite{exp}, 
the beam effects (two-photon correlations) represent a source of systematic errors, which can be estimated knowing the temporal characteristics of the radiation process.

For the kinetic energy of $200$ keV ($\beta = u/c \approx 0.7$), the current $40.8$ nA, and the distance between electrons in the beam of $\Delta z = 0.8$ mm, 
one can estimate \textit{the time of flight} between each electron assuming that the they pass one after another without a transverse shift (see Fig.1, left). 
This assumption per se is incorrect and a real beam rather looks like shown in Fig.1, right. 
This time interval is 
\be
\Delta t = \frac{\Delta z}{\beta c} \approx 4 \, \text{ps}. 
\label{Deltat}
\ee
The probability to emit a photon by each electron on a tree level is roughly $\alpha = 1/137$.
An inverse of the emission rate -- that is, a time interval between events of the photon emission -- is of the order of 
\be
\alpha^{-1} \Delta t \sim 0.5\, \text{ns}.
\label{Deltat2}
\ee
A more realistic estimate must account for the transverse shape of the beam (it decreases this time as seen in Fig.1, right) as well as for details of the specific radiation process. 

\begin{figure}[t]
\centering
\includegraphics[width=.4\linewidth]{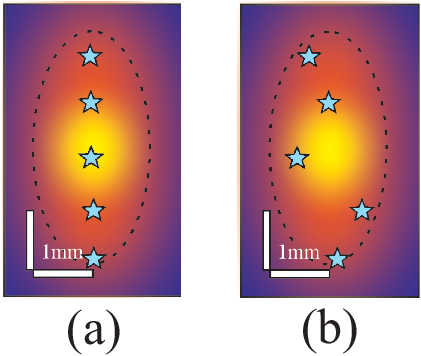}
\caption{Spatial distribution of electrons (marked by stars) in the beam. Left: an idealistic situation to derive the estimate (\ref{Deltat2}). Right: a more realistic picture for which the time $\alpha^{-1} \Delta t$ is smaller.
}
\label{Fig1}
\end{figure}

In order to guarantee that we have a genuine single-electron regime and that the radiation is formed on a width of the transverse coherence length, this time interval must be many orders of magnitude 
larger than \textit{the radiation formation time}, that is, the time during which a virtual photon emitted by the moving electron is scattered by the grating.
Microscopically, the virtual photon excites (polarizes) atoms or conduction electrons on a grating surface, and there is a finite \textit{relaxation time} $\delta t_{rel}$, 
during which the real photon is emitted. Clearly, this time strongly depends on the grating material and on virtuality of the initial photon, i.e. on the electron energy.
A lower bound for this time can be obtained from the uncertainty relation,
\be
\delta t_{rel}\, \delta \omega \geq 1/2.
\label{deltat}
\ee
For the optical wavelengths $\lambda \sim 0.5 - 1\,\mu$m, we have roughly
\be
\delta t_{rel} \geq \frac{\omega}{\delta\omega}\times 1\, \text{fs}.
\label{deltat2}
\ee
A line width of $\delta\omega/\omega \sim 10^{-3} - 10^{-2}$ yields 
\be
\delta t_{rel} \geq 0.1 - 1\, \text{ps}.
\label{deltat2fine}
\ee
For metals, more rigorous estimates based on a finite free path of conduction electrons yield similar numbers.
Unlike Eq.(\ref{Deltat2}), the estimate (\ref{deltat2fine}) is but a lower bound, and the real relaxation time can be much higher.

Thus one can look at the ratio
\be
K_t = \frac{\delta t_{rel}}{\alpha^{-1} \Delta t}
\label{ratio}
\ee
as at \textit{a measure of temporal coherence} of the radiation process generated by the beam. When this ratio approaches unity, the destructive interference of photons emitted by different electrons can well happen
in the pre-wave zone and in this case it is the beam width and not the transverse coherence length that defines the wave zone radius. For parameters of the experiment \cite{exp}, the lower bound for this ratio is
\be
K_t > 10^{-3} - 10^{-2},
\label{ratio2}
\ee
but the exact value is \textit{unknown}. The large samples of data with $N_{\gamma} \gg 1$ points will inevitably contain contributions of two-photon events as well as statistical fluctuations 
with a weight of $1/\sqrt{N_{\gamma}}$. When a sample contains but a few points (like shown in Fig.3 of Ref.\cite{exp}), one can likely neglect the two-photon correlations but the uncertainties of such measurements stay large.
The data in Refs.\cite{exp,com} seem inconclusive to fully rule out the influence of the collective effects because neither the errors nor the sample size were provided.

\section{Other classical effects: the grating shape, its material, the beam divergence}

Another family of classical effects that could lead to broadening of the azimuthal distributions is a role of the beam angular divergence, of the grating shape and of its finite permittivity $\varepsilon (\omega) = \varepsilon^{\prime} + i \varepsilon^{\prime\prime}$. It seems that none of these effects was discussed or taken into account in Refs.\cite{exp, com}. It is only for ultrarelativistic energies with $\gamma \gg 1$ that the grating shape and the angular divergence do not play any significant role and the angular distributions of Smith-Purcell radiation are well described by simple formulas of the surface current models \cite{Kaz, Pap, Br, Black, PLA, Mono} -- see comparison of the different models in \cite{PRAB2006, PLA}. It is so because the coherence length of a virtual photon $\beta\gamma\lambda/2\pi$ is much larger than a size of the grating strip. 

On the contrary, for parameters of the experiment \cite{exp} different models of Smith-Purcell radiation \textit{disagree} and, in particular, they disagree in the predicted width of the azimuthal distributions even for gratings of the same shape -- see Figs.10,11 in Ref.\cite{PRAB2006} (for parameters of Refs.\cite{exp,com} the difference is much larger). So it is generally \textit{not clear} what width of the azimuthal distributions one should expect for these parameters and for the chosen grating \cite{gr} in the wave zone. 

\begin{figure}
\centering
\includegraphics[width=.5\linewidth]{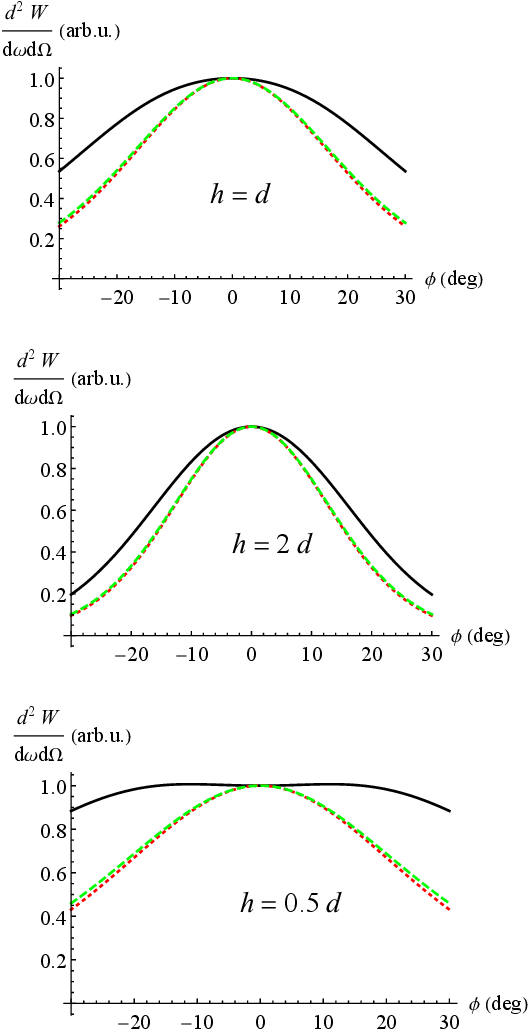}
\caption{The azimuthal distributions of Smith-Purcell radiation by a single electron in the wave zone for different impact parameters $h$ and for the polar angle of radiation $\theta = \pi/2$, the grating period $d = 416$ nm, the wave length $\lambda = 600$ nm, the distance between the strips $d/2$, and the height of each strip $d/4$. The red dotted line -- silver (the permittivity was taken from Ref.\cite{eps}), the green dashed line -- ideal conductor, the black solid line -- a dielectric. The model of Ref.\cite{JETP11} was used. Clearly, if the impact parameters $h \lesssim d$ mostly contribute to the data of Ref.\cite{exp} due to angular divergence of the beam, the azimuthal distributions can be wide thus mimicking the quantum effect of the spatially localized charge. A contribution of the dielectric substrate could lead to the similar effect for large statistics.}
\label{Fig2}
\end{figure}

The width of the azimuthal distributions also depends on the impact parameter and on the grating material. The finite angular divergence of the beam implies that different impact parameters contribute to the measured intensity and in order to judge how this divergence modifies the distributions averaging over the impact parameters -- or, alternatively, over the angles -- is necessary. Finally, it is not clear whether the dielectric substrate of the grating used in the experiment \cite{exp} can contribute to the radiation or not. To illustrate the possible influence, we present in Fig.2 the angular distributions in the wave zone for a grating made of rectangular strips of an arbitrary permittivity $\varepsilon (\omega)$ separated by vacuum gaps \cite{JETP11}. While large impact parameters $h \gg d$ correspond to narrow angular distributions, the values $h \lesssim d$ yield much wider distributions,
especially if there is an influence of the dielectric substrate. Clearly, the contribution of the large impact parameters is exponentially suppressed and the small values of $h$ can be of main importance. To make quantitative estimates, one needs to average over the angles with a form-factor of the real beam taken into account, which was not done in Refs.\cite{exp, com}. The contribution of the dielectric can be neglected only when the metallic surface is \textit{continuous} (it is not clear from \cite{gr, exp} if this is the case), otherwise it is also a source of systematic uncertainties for large statistics.

\section{The role of electron post-selection}

The part 2 of the comment \cite{com} is devoted to explanation that our results in Ref.\cite{PRA} ``only hold for a certain experimental situation where the final state of the emitting electron is observed in coincidence with the emitted photon''. This claim is incorrect because in our paper we studied in detail two scenarios: one in which the final electron is not detected at all (see Sec. III.B) and when the final electron is detected in coincidence with the final photon (see Sec.III.C). We explained the difference and its consequences. The key consequence is that in the former case (the final electron is not detected) the radiation intensity does not depend on a phase $\varphi$ of the initial electron's wave function in the momentum space $\psi ({\bm p}) = |\psi|\exp\{i\varphi\}$, but it still depends on the absolute value $|\psi|$ of this wave function. In other words, the emission rate does not depend on the shape of the electron wave packet (defined by its phase) and it does depend on the size of this packet, defined by the overall envelope in $|\psi|$.
This observation is not in fact a novelty of our work, as it has also been noted in several other papers (for instance in references \cite{HD,KPO}). Moreover, the comment \cite{com} contains 
incorrect statements such as, for instance, ``in the case of paraxial and perturbative free-electron radiation from transversely uniform media, with no electron post-selection, the radiation does not depend on the initial electron wavefunction''. As we have noted, it is known to depend but only on the absolute value of the wave function.

Finally, the authors of the comment note that ``specifying the final electron state as a plane wave ... means that it is in fact detected'' \cite{com}.
While this statement is absolutely correct as it is, we would like to emphasize that choosing the detected state of the final electron as a plane wave can actually describe 
two completely different physical scenarios: 
\begin{itemize}
\item
One in which the final electron \textit{is} detected with an ordinary plane-wave detector, 
\item
And the second one in which this electron is \textit{not} detected at all.
\end{itemize}
This is a well known situation in the theory of neutrino oscillations (see for instance Ref.\cite{Akh}), in which the undetected particle's wave packet stays completely \textit{delocalized} in space.
This corresponds to a plane wave with the momentum simply following from the energy-momentum conservation law, provided that the momentum of the final photon (in our problem) is measured 
and the recoil of the grating can be neglected.

\section{Conclusion}

We have argued that while the main problem, discussed in the comment \cite{com}, of whether the measurements of Ref.\cite{exp} were performed in the wave zone is interesting and worth scrutiny, 
it is of secondary importance for the key question, which is whether or not those measurements allow one to conclude in favor of the localized-charge hypothesis.
In our view, they do not and the observed lack of dependence on the electron's transverse coherence could have been expected from the general considerations,
provided that the electron packets are at least approximately Gaussian \cite{PRA}. 

On the other hand, the unusually large width of the azimuthal distributions reported in Ref.\cite{exp} remains unexplained in the comment.
As we have showed, there is a family of classical effects that could also lead to the wide azimuthal distributions and that have neither been taken into account nor even mentioned in the paper \cite{exp} 
and in the comment \cite{com}. These effects represent alternative hypotheses that must be carefully checked in order to make an unambiguous conclusion in favor of one of the hypotheses. 

Summarizing, we think that the choice of Smith-Purcell radiation for studying the influence of the electron packet's width on the radiation characteristics is an unfortunate one,
because there are too many subtle effects that must be taken into account to make reliable conclusions. A much better candidate for such measurements seems to be \textit{diffraction radiation} from a thin conducting semi-plane 
or a rectangular strip. The radiation formation length is much shorter in this case and the beam divergence and the temporal coherence of the radiation process do not play such an important role. In addition, the theory of such a process (see, for instance, \cite{PLA, JETP11, Mono}) leaves less room for controversy than it is for a more complex geometry of Smith-Purcell radiation. Thus the results of such measurements would have higher credibility. Another candidate is emission in external electromagnetic fields (for instance, a Compton emission in laser pulse) because in this case it is the electron itself that emits a photon and so the transverse coherence length is expected to play a major role.

\

We thank A.A. Tishchenko for useful discussions. This study is supported by the Government of the Russian Federation through the ITMO Fellowship and Professorship Program.

\end{document}